\documentclass[preprint,12pt]{elsarticle}

\usepackage{lscape}
\usepackage{graphicx}
\usepackage{amssymb}
\usepackage{amsthm}
\usepackage{amsmath}
\usepackage{makecell}
\usepackage{setspace}
\usepackage{geometry}
\usepackage{lscape}
\usepackage[hidelinks=true]{hyperref} 
\usepackage{soul}
\usepackage{extract}

\usepackage{lineno}

\DeclareMathOperator{\Tr}{Tr} 

\usepackage{xcolor}
\newcommand{\cinl}[1]{} 
\renewcommand{\cinl}[1]{\textcolor{olive}{\textsc{{#1}---CSM}}} 

\newcommand{\jinl}[1]{} 
\renewcommand{\jinl}[1]{\textcolor{blue!40}{\textsc{{#1}---Jeff}}} 

\newcommand{\finl}[1]{} 
\renewcommand{\finl}[1]{\textcolor{red!40}{\textsc{{#1}---Felix}}} 

\newcommand{\linl}[1]{} 
\renewcommand{\linl}[1]{\textcolor{magenta!40}{\textsc{{#1}---Laura}}} 

\journal{XXXX}

\begin{document}

\begin{frontmatter}


\title{An efficient counting method for the colored triad census}



\author[nih,ox,cor]{Jeffrey Lienert}
\author[nih]{Laura Koehly}
\author[ox]{Felix Reed-Tsochas}
\author[nih,cor]{Christopher Steven Marcum}

\address[nih]{Social Network Methods Section, National Human Genome Research Institute, National Institutes of Health}
\address[ox]{CABDyN Complexity Centre, Sa\"{i}d Business School, University of Oxford}
\address[cor]{Corresponding Author}

\begin{abstract}
The triad census is an important approach to understand local structure in network science, providing comprehensive assessments of the observed relational configurations between triples of actors in a network. However, researchers are often interested in combinations of relational and categorical nodal attributes.  In this case, it is desirable to account for the label, or color, of the nodes in the triad census.  In this paper, we describe an efficient algorithm for constructing the colored triad census, based, in part, on existing methods for the classic triad census.  We evaluate the performance of the algorithm using empirical and simulated data for both undirected and directed graphs.  The results of the simulation demonstrate that the proposed algorithm reduces computational time many-fold over the na{\"i}ve approach. We also apply the colored triad census to the Zachary karate club network dataset.  We simultaneously show the efficiency of the algorithm, and a way to conduct a statistical test on the census by forming a null distribution from $1,000$ realizations of a mixing-matrix conditioned graph and comparing the observed colored triad counts to the expected. From this, we demonstrate the method's utility in our discussion of results about homophily, heterophily, and bridging, simultaneously gained via the colored triad census.  In sum, the proposed algorithm for the colored triad census brings novel utility to social network analysis in an efficient package.

\end{abstract}


\end{frontmatter}



\doublespace

\section{Introduction \label{S:1}}

The triad census is an important approach towards understanding local network structure. \citet{Holland1976} first presented the 16 isomorphism classes of structurally unique triads possible in a directed network without loops.  To conduct a triad census, one simply counts each occurrence of these structures, without respect to the labeling of the nodes (here we use node label, color, characteristic, and attribute interchangeably).  This is useful insofar as specific triads, or combinations thereof, may relate to underlying social processes giving rise to an observed network.  For example, bridges (triads with one null dyad and two non-null dyads) may be important in navigating social networks \citep{Granovetter1973}, and certain triads may be more or less favorable based on structural balance theory (e.g. the 300 is balanced but the 201 is not, see Figure ~\ref{fig.tc}) \citep{Cartwright1956}.  Moreover, a variant of the triad census, motif analysis, investigates the statistics of various triad configurations (motifs), and has found wide application in biology \citep{milo2002network}.  \par

Also important to network structure are nodal characteristics and how they relate to tie formation or dissolution.  This has been the subject of research on homophily (individuals having similar attributes with those to whom they are connected) \citep{McPherson2014}. However, homophily is an observed phenomenon, not a process. The processes giving rise to homophily are varied, often confound the relationship between networks and outcomes, and are difficult to tease apart \citep{Shalizi2012}. Methodological advances, such as stochastic actor-oriented models can disentangle these effects to some extent \citep{snijders1996stochastic}. Other analyses have attempted to disentangle the processes leading to homophily from structural processes, such as triadic closure \citep{goodreau&2009bfff}. Additionally, the coloring of nodes in a network has been an important question for many graph theorists and indeed represents a major topic in this field \citep{Jensen2011}. \par
    
Although nodal characteristics and the triad census are important, they have rarely been examined fully in conjunction. Yet, there are a few cases where specific colored triads have been studied.  For example, \citet{gould1989structures} study brokerage based on triad structure and group membership simultaneously.  This same approach has been used to study brokerage in dynamic networks \citep{SPIRO2013130}.  As well, a study by \citet{Marcum2015} examined specific colored triads based on generational membership within families; in this work the authors showed that inter-generational ties were observed in different quantities than expected based on the underlying null model.  None of the past research evaluated the full census of colored triads, rather, researchers have focused instead on specific colored triads that were \emph{a priori} expected to be relevant to the processes at hand.  As a result, these foundational works were not exhaustive with respect to all alternatives. In other words, previous research examining a subset of colored triads likely had an amount of false negatives due to not examining every colored triad; this could be addressed by censusing the colored triads.  \par

The examination of node characteristics together with local structure is important as it provides opportunity to simultaneously study the occurrence of triadic structure, nodal attributes, and the interactions between them.  For instance, certain colored triads may be forbidden, such as three-cycles between strict heterosexuals in mixed-orientation sexual contact networks \citep{marcum2016fourcycle}.  Impermissible triads would be categorized the same as those that were not observed due to chance in a triad census, potentially missing important social processes or constraints at play in this type of network.  Only by incorporating node coloring into the triad census can this pattern be fully elucidated. \par
    
Based on this methodological gap in the literature, we develop a method to census the colored triads for any one-mode binary network with arbitrary number of colors. Due to the large numbers of isomorphism classes of size 3 as the number of colors increases, this method requires computational efficiency in addition to mathematical accuracy. As well, one is often interested in forming a null distribution with which to compare observed colored triad counts. If the null distribution cannot be analytically solved, one would likely census the colored triads of many simulated networks, further increasing the need for the algorithm to be computationally efficient. \par

Current efficient methods for the triad census exploit the sparseness of networks \citep{Batagelj2001}, and scale sub-quadratically (as the number of edges increases the time to run the algorithm is faster than the number of edges squared).  However, methods that exploit network sparseness by inferring the number of null triads do not work in the colored case because they do not explicitly interrogate every triad, and there are variations within the null triads due to the coloring.  Therefore, we extend the methodology of \citet{Moody1998}, which is based on matrix algebra and interrogates every triad; his method scales sub-quadratically with the number of nodes. \par

This paper (1) presents the colored triad census and its computational complexity, (2) shows that this approach can be used on large networks (tested for up to $10,000$ nodes) with up to 10 colors in relatively efficient time, and (3) uses the method many times to create null distributions of colored triad censuses to form the basis of conditional uniform graph tests. We illustrate the benefits of an analysis incorporating the colored triad census using a well-known dataset, Zachary's Karate Club \citep{zachary1977information}. \par

\section{Algorithm \label{sec.alg}}

Since the original appearance of the triad census in 1976, a number of papers have explored how to compute the triad census of a network in an efficient manner.  Although, for sparse networks, sub-quadratic methods (in terms of number of nodes) exist for calculating the triad census (e.g. \citet{Batagelj2001}), we use the quadratic algorithm presented by \citet{Moody1998} here. This is because the more efficient methods avoid interrogating null triads directly by taking advantage of the sparseness of graphs, the subsequent large number of null (003) triads, and the known number of total triads.  Instead, they interrogate all triads with at least one edge, and then subtract that count from the total number of triads in the network to arrive at the number of null triads.  This is insufficient in the colored triad census as there are differently-colored null triads, and the count of each cannot therefore be algebraically determined. For example, if there are two colors, four different null colored triads exist (0-3 nodes color A). The exact breakdown of the null triad into the four colored triads cannot be determined without interrogating each null triad, thereby losing the efficiency gained when not considering colors.  Moody's algorithm does not employ this limiting shortcut, and we therefore use it as a basis for our colored triad census algorithm.  Additionally, because many networks are sparse, we can leverage computational techniques for increasing the efficiency of sparse matrix operations \citep{duff2017direct}, further reducing the computational complexity of our method.

\citet{Moody1998} showed that the count of each of the 16 triad isomorphism classes could be derived by using matrix algebra on the adjacency matrix of the graph and its derivatives.  To review, let $\mathbf{A}$ be the adjacency matrix of a network, and $A_{ij}=1$ when a tie exists from node $i$ to node $j$. Let $E$ be the symmetrized matrix $A$, formed by making any edge in $A$ reciprocal via $E_{ij}=\max(A_{ij},A_{ji})$. The complement of $E$, $\bar{E}$, is formed by subtracting the complete network adjacency matrix from $E$, so that $E_{ij}=1$ if and only if there is neither a tie from $i$ to $j$ nor a tie from $j$ to $i$.  Next, we have $M$, the mutual matrix of $A$, and is made by removing any asymmetric edges from $A$, or $M_{ij} = \min(A_{ij},A_{ji}) $. Finally, $C$ is the matrix of only asymmetric edges, and is calculated by $C=A-M$. Therefore, $C_{ij}=1 \iff A_{ij}=1 \; \& \;  A_{ji}=0$. Based on these matrices, Moody demonstrates how to calculate the number of each of the 16 isomorphism classes for the case of unlabeled graphs (or, equivalently, for a graph consisting of nodes of the same single color). Generally, this was done by multiplying (either through dot-product or element-wise multiplication) the three matrices corresponding to the relevant edges in the triad of interest. There were two triads ($111U$ and $111D$) that were not directly amenable to this process and were calculated via addition and subtraction of other triad types, respectively.  \par

To extend this work to the case of multiple colors, we introduce the out-coloring and in-coloring matrices, $K^{r}$ and $K^{r'}$, respectively, where $r$ is the focal color of matrix $K$.  Here, the in-coloring matrix is the transpose of the out-coloring matrix.  The out-coloring matrix is calculated by evaluating the color of the nodes row-wise, such that rows indexing nodes of the focal color are composed in the following way:
\begin{equation}\label{eq.kr}
	K^{r}_{i\bullet} =
	\begin{cases}
    	1   &   \text{if } R(i) = r	\\
    	0  	&	\text{if } R(i) \ne r 
  	\end{cases}
\end{equation}
Where $R(i)$ is a function returning the color of node $i$.  As above, the in-coloring matrix is the transpose of the out-coloring matrix in Eq.~\ref{eq.kr}. \par

\begin{figure}[ht]
\centering\includegraphics[width=1.0\linewidth]{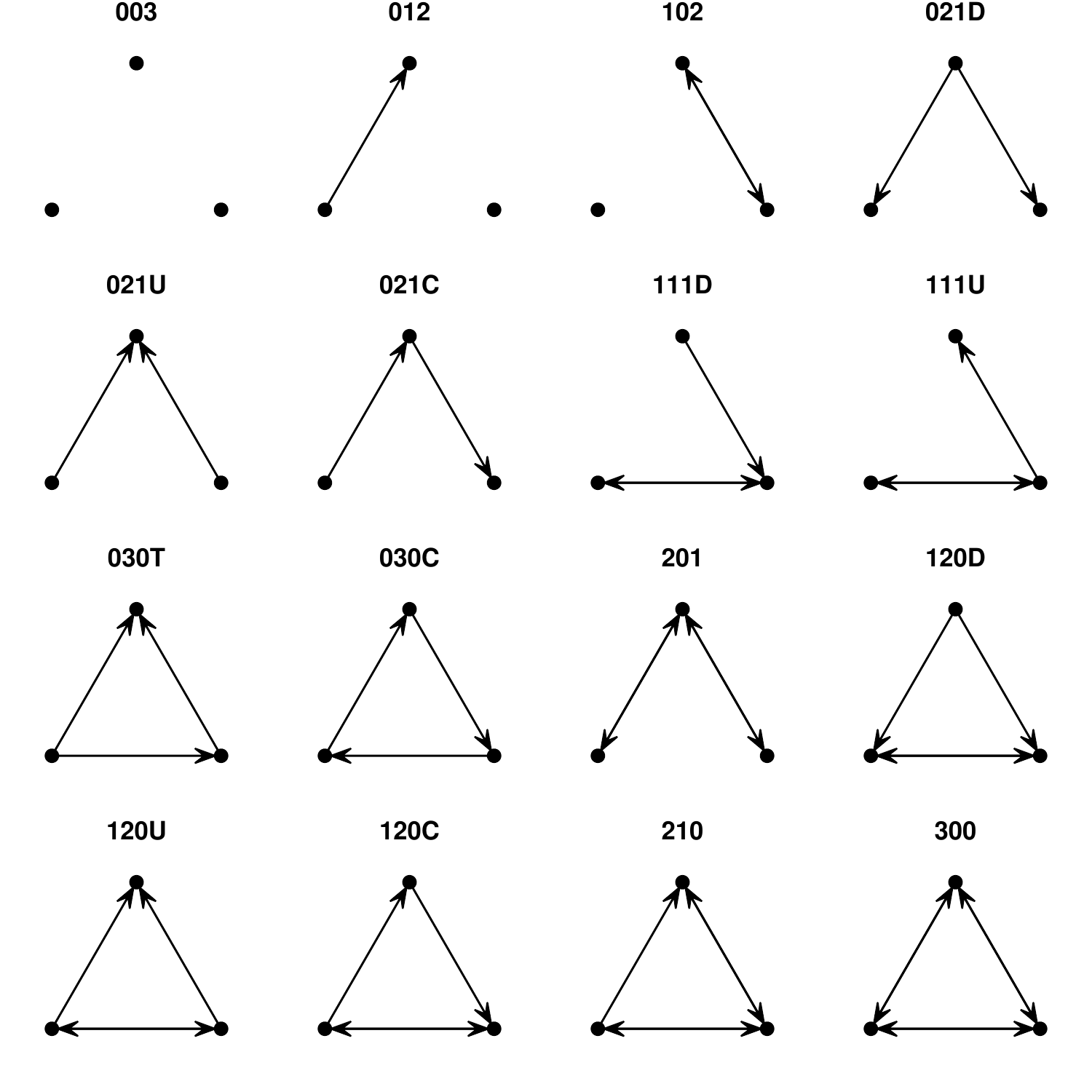}
\caption{The 16 isomorphism classes of triads and their orientation used here with respect to the color numbering.  When colors are added to these triads, they are labeled starting from the top node and proceeding clockwise. \label{fig.tc}}
\end{figure}

Our algorithm works by using the in- and out-coloring matrices to evaluate and ``switch on" edges that have nodes of the focal colors at the \textit{ends} (or tails) of edges in the adjacency matrix $A$ of the network.  We adapt the triad census nomenclature of \citet{Holland1976} by appending the colors after the name of the triad. The colors are ordered from the top node proceeding clockwise in Figure ~\ref{fig.tc}.  We have arbitrarily adapted the orientation of the triads from the triad census figure in \citet{Holland1976} for computational reasons.  The orientation is important here because triads with the same orientation may no longer be isomorphic when color is introduced.  Figure~\ref{fig.tc} makes it possible to  count unambiguously and name only unique colored triads.  Therefore, $T_{102-123}$ is the triad consisting of 1 symmetric dyad and 2 null dyads, where the top node is of color $1$, the bottom-right node is of color $2$ and the bottom-left node is of color $3$. This is distinct from the $T_{102-312}$ triad because the coloring of the nodes is not identical from the previous triad. \par

Following this, the general formula for an arbitrary triad ``\textit{T}" with an arbitrary coloring triplet is:

\begin{equation}
T=\Tr\big((K^{1}  \times H(T,1,2) \times K^{2'})(K^{2}  \times H(T,2,3) \times K^{3'})  (K^3 \times H(T,3,1) \times K^{1'})\big) 
\end{equation}  

In the above, ``$\times$" refers to element-wise multiplication, and ``Tr" is the trace function. For an arbitrary triad, ``\textit{T}" has a color triplet $r_1,r_2,r_3$. $H(T,i,j)$ is a function returning the matrix specific to the type of edge between nodes $i$ and $j$ in triad ``$T$".  For example, in a $102$ triad, the first edge from the top node going clockwise is a symmetric edge from node one to node two (Figure~\ref{fig.tc}). $H(T_{102},1,2)$ in this case would be the matrix $E$ for the symmetric matrix, and the sandwiching color matrices would turn the proper edges on and off if nodes one and two were of the specified colors.  If the edge is an asymmetric one, and the direction of the edge in Figure \ref{fig.tc} is counter-clockwise, then $C'$ is used instead of $C$ to force the edge to go in the proper direction.

At this point, there are redundant triads due to certain colored triads being isomorphic. For instance, the $T_{003-122}$ is isomorphic with $T_{003-221}$ and $T_{003-212}$, and would be triple-counted. These are removed by checking for isomorphisms based on matrix row and column permutations of the triad. If two colored matrices are identical after such row and/or column permutations, then they are isomorphic, and one is removed. We arbitrarily decide to discard the triad whose coloring triplet name comes second alphanumerically. It should be noted that removing in this way is computationally expensive, particularly as the number of colors and nodes grows large.  We therefore shorten this process by performing it once for 1 to 10 colors and storing the unique isomorphism classes. This leaves only unique isomorphism classes of colored triads, which can then be accessed in linear time. 

The number of unique isomophism classes for a given number of colors can be shown for each of the 16 ismorphism classes in the triad census. The 16 classes separate into four types of colored triads, depending on how many structurally-distinct positions there are in the triad (e.g. the two ends of the edge in a 102 triad are not structurally-distinct from one another, but are distinct from the node with no edges).  The calculation for the number of each isomorphism class for arbitrary number of colors ($k$) is shown in Table~\ref{table.ct}.  Each combinatoric term in each row (together with their respective leading permutation coefficients) counts the number of colored triads when there are three, two, or one unique color(s), respectively. For example, in a network with three colors,the `300' and `003' classes have only one accessible permutation when there are three colors present in the triad (i.e. $\binom{3}{3}$), six ways when there are two colors (i.e. $2\,\binom{3}{2}$), and one way when there is one color in the triad (i.e. $\binom{3}{1}$).
\begin{table}[ht]
\centering
\begin{tabular}{c c}
\hline
\textbf{Isomorphism classes} & \textbf{Number of colored triads} \\[5pt]
\hline
$300$ and $003$ & \, ${{k}\choose{3}}+2{{k}\choose{2}}+{{k}\choose{1}}$ \\[5pt]
$030C$ &  $2{{k}\choose{3}}+2{{k}\choose{2}}+{{k}\choose{1}}$ \\[5pt]
$102$ $021D$ $021U$ $201$ $120D$ and $120U$ &  $3{{k}\choose{3}}+4{{k}\choose{2}}+{{k}\choose{1}}$ \\[5pt]
$012$ $021C$ $111D$ $111U$ $030T$ $120C$ and $210$ &  $6{{k}\choose{3}}+6{{k}\choose{2}}+{{k}\choose{1}}$ \\[5pt]
\hline
\end{tabular}
\caption{Expression for the number of isomorphism classes within a triad class. $k$ is the number of colors \label{table.ct}} 
\end{table}

If these numbers are summed over the 16 isomorphism classes, the total number of colored isomorphism classes of triads for $k$ colors is returned.  Similarly, the same can be done for undirected triads, solely summing over the $4$ triads observed in the undirected case.  Table~\ref{table.count} reports the total number of colored triads for undirected and directed networks over a range of $k$. Clearly, the number of isomorphism classes grows quite quickly as $k$ increases.

\begin{table}[ht]
\centering
\begin{tabular}{c c c}
\hline
\thead{Number of \\ colors} & \thead{Number of directed \\ colored triads} & \thead{Number of undirected \\ colored triads}\\
\hline
1 & 16  & 4 \\ 
2 & 104 & 20\\ 
3 & 328 & 56\\ 
4 & 752 & 120\\ 
5 & 1440 & 220\\ 
6 & 2456 & 364\\ 
7 & 3864 & 560\\ 
8 & 5728 & 816\\ 
9 & 8112 & 1140\\ 
10 & 11080 & 1540\\ 

\hline
\end{tabular}
\caption{The number of colored triad isomorphism classes for directed and undirected networks for $k$ ranging from 1 to 10. \label{table.count}}
\end{table} 

The algorithm implemented as an R package is publicly available and is linked to this paper via github: \url{https://github.com/jlienert/ColoredTriadCensus}. \par

\section{Algorithmic Performance}

If a na{\"i}ve implementation of matrix multiplication is used, this algorithm runs with computational complexity $O(N^3*3^k)$. It scales with the number of nodes cubed ($N^3$) because of the matrix multiplication involved in the algorithm. However, many software packages use algorithms that reduce the complexity of matrix multiplication to $O(N^{2.38})$ \citep{davie_stothers_2013}. Furthermore, by taking advantage of methods for matrix multiplication using sparse matrices (as appropriate due to the sparse nature of most social networks), this complexity is reduced to something closer to $O(N^2)$ \citep{yuster2005fast}. The exact benefit gained by using sparse matrix multiplication varies based on how sparse the matrix is. This ranges from the nearly-optimal $O(N^2)$ when very few edges exist, to worse than the optimized algorithm when many edges exist. The scaling with $3^k$ comes from the number of distinct colored triads the algorithm needs to evaluate, and the number of isomorphism classes scales in such a manner. \par
	
\begin{figure}[ht]
\centering\includegraphics[width=1.0\linewidth]{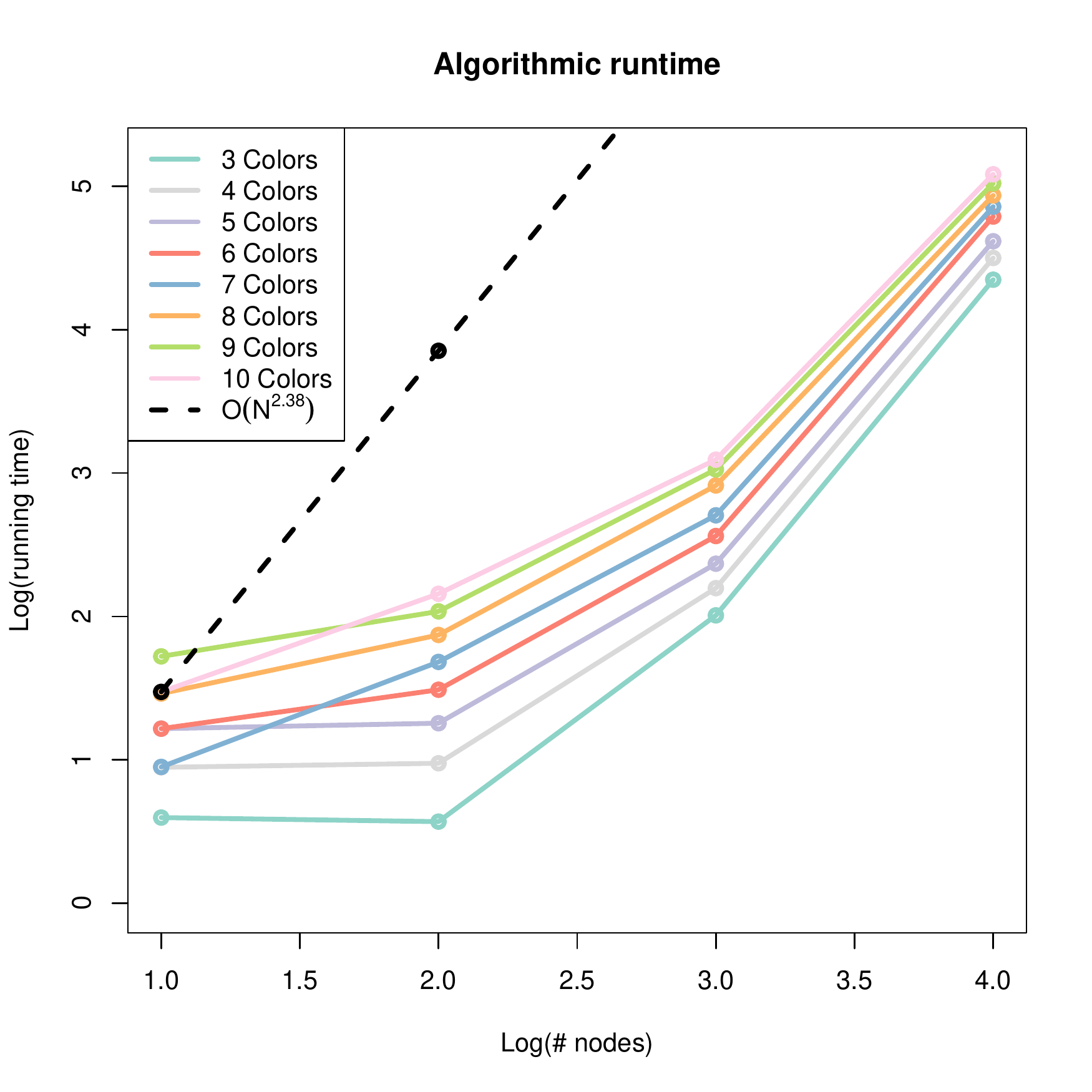}
\caption{Runtime of the algorithm on networks ranging from size $10$ to $10,000$ nodes in orders of magnitude, and from one to ten colors. Additionally, the dashed line represents the computational time that would be expected using standard matrix multiplication methods for $k=10$. These runtimes were generated using virtual PCs including 1 dual core CPU and 10GBs of RAM. \label{fig.rt}}
\end{figure}

To test the efficiency of the algorithm, we apply it to networks ranging in size from $n=10$ to $n=10,000$ with the number of colors ranging from $k=3$ to $k=10$, all holding the average degree constant at 6 by creating Erd\"os-R\'enyi graphs with an edge probability of $\frac{6}{N-1}$. This reflects the average number of ties participants enumerate in social networks surveys \citep{marsden2003interviewer}. The runtime of the algorithm with these parameters can be seen in Figure~\ref{fig.rt}.  In general, increasing $K$ results in constant increases in $\log(runtime)$, which is what we expect based on the theoretical computational complexity.  As expected, we also observe a super-quadratic increase in $\log(runtime)$ as $N$ increases. Although it is super-linear, it is still below the curve that would exist if we used matrix multiplication not optimized for sparse matrices (dotted line in Figure ~\ref{fig.rt}). This difference shows the expected time saved by using sparse matrix methods.  Finally, we observe changes in the rank-order and decreases in runtime going from $10$ to $100$ nodes. This is also due to the computational time involved in initializing the sparse matrices and storing and operating on sparse matrices, and as such is not unexpected. Additionally, because the average degree was held constant, the smaller networks are much more dense, and therefore are actually less efficient than if they used standard matrix multiplication methods. To be perfectly optimized, therefore, the algorithm would use standard matrix multiplication for small networks, and switch to  sparse methods for larger networks. However, the gains would be minimal, generally under 10 seconds, and would require additional logical steps to check for network size, further minimizing the gain. We therefore use sparse matrix methods for all network sizes. \par
    
\section{Empirical Use and Example}

To show the empirical value of this algorithm, we use the Zachary karate club social network \citep{zachary1977information}. This is a well-known historical network that describes the social relationships between 34 members of a university karate club. Ties exist between members if they overlapped in at least one of eight contexts representing undirected relations. These relations varied in terms of likely strength of the association. Likely at the weak end of the spectrum is being enrolled in the same class at the university, while likely at the strong end is being a student-teacher at the studio. Additionally, three ties are specific to activities with a part-time instructor.  Member ``factions" were identified as a node attribute, taking one of five mutually exclusive values: strongly associated with the president, weakly associated with the president, neutral, weakly associated with the part-time instructor, or strongly associated with the part-time instructor. These are labeled "Zs", "Zw", "N", "Hw", and "Hs", respectively. These labels can be placed on an ordinal scale from -2 (Zs) to 2 (Hs) to quantify members' direction and strength of alignment. This undirected network with five colors represents a case that is rich in the number of colored triads (220) for detailed conclusions to be drawn using the proposed algorithm (which is general to both undirected and directed networks). \par
    
We initially ran the colored triad census on the social network using the faction as the nodal attribute. This provided the basis for our empirical observed colored triad census. To determine whether these triads were observed more or less often than expected by chance, we constructed a null model. As the choice of null model can have important ramifications for the null distribution of triads, we chose a model where edge formation is a function of the probability of ties between nodes of specific attributes \citep{faust2010puzzle}. The null model is a mixing-matrix conditioned uniform random graph distribution based on probabilities of edges between nodes of particular color combinations \citep{newman2003mixing}. This matrix comprises empirical probabilities of ties between groups, with the diagonal representing within-group tie probabilities. Observations of significantly over- or under-represented colored triads are the result of network effects beyond homophily and heterophily. Networks are then generated from this matrix via a Bernoulli random graph process \'a la \citet{ER1959}. This null model therefore conditions on graph size, the distribution of node factions, and the probability of ties within and between factions.  By generating networks from the null model, we can observe whether colored triad counts deviate from that expected based on the marginal distribution of faction mixing.  Because we condition on the above parameters, if we observe statistical deviations in our colored triad census, it indicates that the structure of the network is dependent on parameters other than those on which we conditioned.  \par	
    
Moreover, for any triad, the expected number and variance can be calculated assuming each tie follows a Binomial distribution (which is a reasonable assumption for most binary social network data). The observed number can then be compared to these numerical results and a p-value extracted from an exact Binomial test. This equates to the following probability, expectation, and variance for an example colored triad:
    
    \begin{align}\label{eq.pt}
    \begin{split}
    P(T) =& P(A_{ij}=1|R(i)=r_1 , R(j)=r_2) \times P(A_{ij}=1|R(i)=r_2 , R(j)=r_3) \\
         \times & P(A_{ij}=1|R(i)=r_3 , R(j)=r_1) 
    \end{split}    
    \end{align}
    \begin{equation}\label{eq.et}
    E(T) = P(T) \times \prod_{r=1}^{L(T)}{{\sum K^r_{\bullet1}}\choose{S(T,r)}}
    \end{equation}
    \begin{equation}\label{eq.vt}
    V(T) = E(T) \times (1-P(T))
    \end{equation}

The probability of $T$, $P(T)$ in Equation~\ref{eq.pt} is based on the mixing-matrix of the three colors ($r$) involved in the triad $T$.  As is standard for the mixing-matrix approach, this continues to assume that all edges in the graph are independent.  For the expected value of a specific triad, we multiply the probability of a single one of those triads by the total number of colored triplets that exist in the graph. In Equation~\ref{eq.et}, the expectation of the triad, $L(T)$ returns the number of unique colors in $T$ and $\sum K^r_{\bullet1}$ is the number of nodes of color $r$ in the graph.  Also, we take the nodes one, two, or three at a time depending on how many times that color repeats in $T$, represented by $S(T,r)$. This expectation therefore follows a binomial distribution, and it's variance follows accordingly in Equation~\ref{eq.vt}.\par

However, to show that this method also works for null distributions that are not analytically solvable, we construct a null distribution based on simulated draws from the null model. As the number of trials increases, the simulated null distribution of the colored triad census should asymptotically approach the analytical solution shown above.  For each of $1,000$ trials, we draw random networks from the null distribution, and run the triad census on all these networks.  Comparing our observed count to the null distribution then allows us to get an approximate p-value for a conditional uniform graph test, and test the over- or under-representation of each colored triad.  We now turn to these results. \par

\subsection{Results}
\begin{landscape}

    \begin{figure}[ht]
\centering\includegraphics[width=1.0\linewidth]{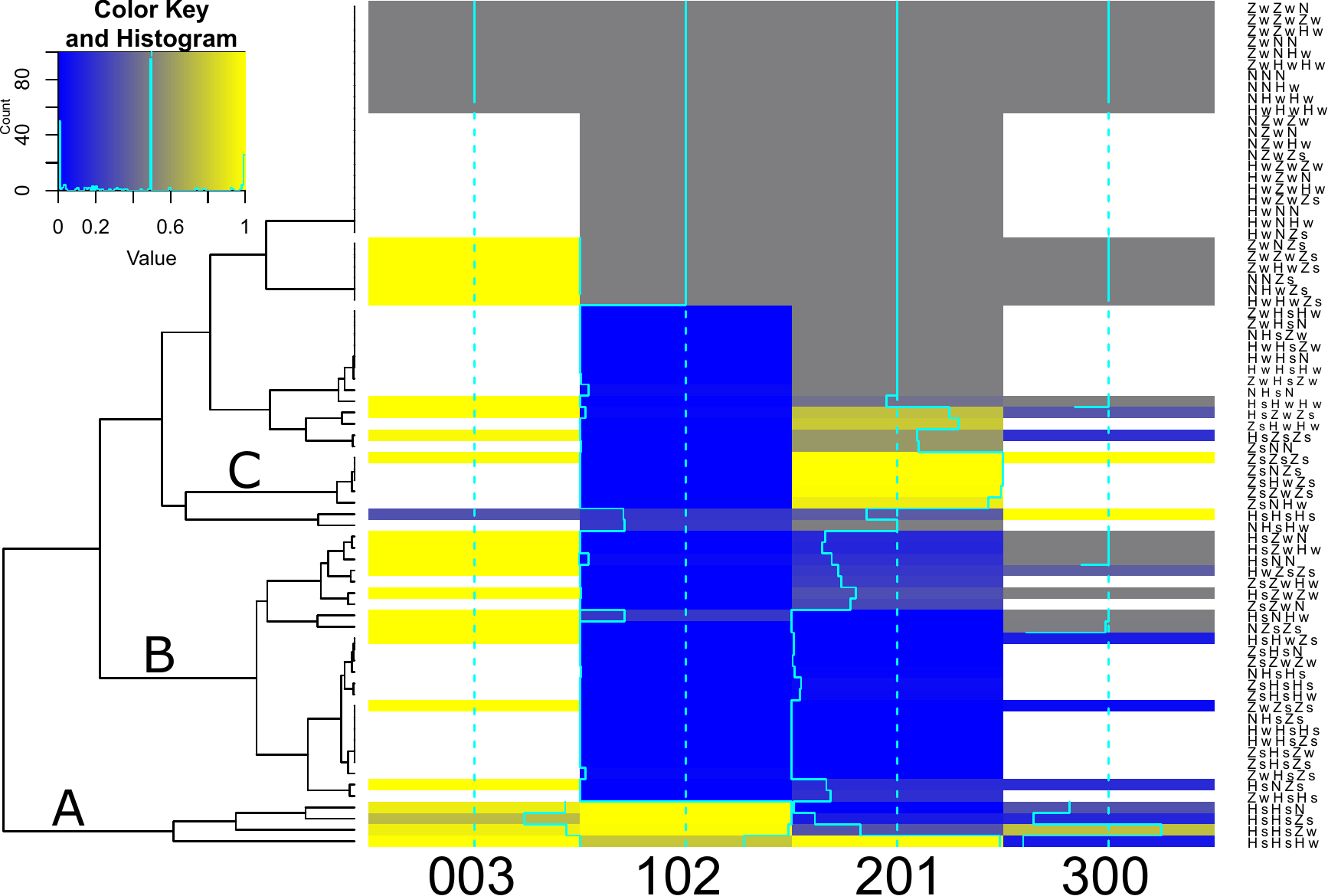}
\caption{Heatmap of colored triads and their corresponding p-value of how often they were observed in the empirical networks relative to the null distribution. The columns separate triads based on the MAN configuration, and the rows separate triads based on the triplet of colors. Standard clustering algorithms were used to create the dendrograms. White space indicates redundant isomorphism classes. Gray boxes are either those with 0 triads observed in the network or in any of the networks of the null distribution, and therefore have an undefined pseudo p-value, or those with a pseudo p-value of $0.5$. The three labels correspond to three breakpoints in the clustering that separate meaningful groups. (A) is a group of four color triplets exhibiting homophily between $Hs$ nodes. (B) is a group of 21 colored triplets exhibiting low clustering between heterogeneous nodes. (C) is a group of 6 colored triplets that show potential significant amounts of bridging.  \label{fig.hm}}
\end{figure}

\end{landscape}

Figure ~\ref{fig.hm} is a heatmap of the approximate p-values associated with each binomial exact test against the null for each triad, clustered by the triad and the colored triplet as returned by the proposed algorithm.  We use a clustering algorithm to group color triplets with similar profiles across the types of triads.  This assists with identifying trends across different colored triads, leading to conclusions that would likely be missed if all the colored triads were individually examined.  We find particular importance in three branch cutpoints in the clustering algorithm on the color triplets.  The first branch in the clustering algorithm (A in Figure~\ref{fig.hm}) separates four color triplets, comprising 16 colored triads, with a pattern of over-observed 003 and 102 triads, and under-observed 201 and 300 triads. These results show that these color triplets are those that are less clustered than expected by chance. The color triplets all contain nodes of two factions with the first two nodes being $Hs$, that is, those strongly aligned with the part-time instructor.  This indicates that those who are so aligned are likely to form ties to one another, but not to members of other factions. The only exception in this group is that two $Hs$ nodes are more likely to form a tie from one of the $Hs$ members to a $Hw$ member, but even in this case the complete triad (003) is still observed less than expected by chance. This particular result is, perhaps, unsurprising, since $Hs$ and $Hw$ members are close in alignment, more so than with those aligning with the president. Therefore, given the tendency towards homophily they are likely to overlap, though less strongly than members of the same faction; hence, the under-observed $T_{003-HsHsHw}$.  \par
    
        \newgeometry{left=1.5cm,right=1.5cm,top=1.5cm,bottom=1.5cm}
\restoregeometry
\doublespacing
    
The second branching point in the clustering (B in Figure~\ref{fig.hm}) separates the group of color triplets that are over-observed for the 003 triad, under-observed for the 102 and 201 triads, and observed about as much as expected for the 300 triads. All the triplets in question have nodes of different factions in the first and second position. Because the edge in the 102 triad is between the first and second node in the triplet (Figure~\ref{fig.tc}), this means that these are all triplets where the first edge is less likely than expected by chance, and the lack of formation of the first edge subsequently hampers the formation of the edge between the second and third nodes in the triplet (201 triad). The first two nodes of these triplets are often (e.g., 16 out of 21) two factions at least a distance of two away (e.g. $N$ and $Hs$), indicating members of a faction are not likely to overlap with members who are too disparate from their faction.  Put another way, this pattern of triads shows a lack of faction heterophily. \par
    
The third branch point (unlabeled) is primarily singling out the group of color triplets that were not observed in the network, and we cannot draw conclusions about their prevalence.  The fourth branch point (C in Figure~\ref{fig.hm}), however, distinguishes a group of five triplets that are under-observed for the 102 triad and over-observed for the 201 triad. This means that the edge between the first two nodes is less likely than expected by chance, but once that edge does occur, the second edge occurs more often than expected by chance. All these triplets begin with a $Zs$ member, and the 201 triad in this case is effectively a bridging tie between it and another. Interestingly, the bridging node is anything other than an $Hs$ (whom are primarily consigned to this role in branch $A$, as discussed above). The third node was another $Zs$ member in four of five triplets. This indicates that $Zs$ members of the karate club did not often overlap members of other factions, but when they did, provided it was not with an $Hs$, that second person also often overlapped with another $Zs$. \par
    
Although the above examples show homophily and bridging, analyzing the full colored triad census allows us to draw further conclusions by looking at other colored triads. In particular, the homophily has mostly been a story of the $Hs$ nodes, and the bridging primarily about the $Zs$ nodes. The 300 triad of both of these factions, when comprising three nodes of the same faction, are observed more often than expected by chance in both cases, which has different implications on the previously-noted results. For the $Hs$ nodes, homophily is strengthened, as not only do $Hs$ nodes not often overlap with members of other factions, they also very strongly overlap with one another. This may partially be an artifact of the types of overlap, as stated before, three of the overlap activities involve direct participation in the part-time instructor's studio, but there are no corresponding groups for the president. This means that those who are $Hs$ or $Hw$ may have more opportunity to overlap with one another due solely to the structure of the data. On the other hand, the triplet of all $Zs$ members also has an over-observed 300 triad. Although there are other triads that seem to indicate bridging between $Zs$ members (C in Figure~\ref{fig.hm}), given that $Zs$ members are also densely connected to one-another, the practical effect of these potential bridging ties is reduced. Observing this joint effect of homophily and bridging ties was possible only through the complete colored triad census. Neither a standard triad census nor a brokerage analysis would have revealed the intricacies of these results. \par
    
In sum, it is clear from these results that the colored triad census allows one to examine multiple trends simultaneously that are often done in isolated analyses, including homophily, heterophily, and brokerage.  Importantly, it also allows for generalizations based on the clustering of various triads or color triplets, as well as specific results based on individual triads. In this manner, the colored triad census can yield results on multiple structural levels simultaneously, all while examining local structure, nodal attributes, and their interaction---that is, net of all alternatives involving mixtures of node coloring and triadic configurations. \par
    
\section{Limitations}

There are some limitations to this method.  First, it is only computationally efficient relative to existing methods (including brute force counting).  Networks of $10,000$ nodes or more will take over a day to run using the proposed algorithm for the colored triad census. However, this is an easily parallelizable process (by partitioning the separate algebraic steps, for example), and so the real time necessary to run the analysis can be greatly reduced by taking advantage of this feature.  The time needed for the parallelized colored triad census is approximately inversely proportional to the number of computational cores used in the calculation (plus some overhead).  Second, the interpretation and visualization of these results is complicated, particularly as the number of colors increases.  Examining all of the triads simultaneously reduces the likelihood of missing interesting results because a specific colored triad was excluded.  However, the sheer number of colored triads means that making complete sense of results can be difficult.  Even if the results are carefully examined for all colored triads, it is conceivable that one might miss an important result out of the $11,080$ colored triads in a directed, 10-color network, no matter how meticulous the examiner's eye. However, use of standard clustering algorithms and heatmaps (as in Figure ~\ref{fig.hm}) may help to ease interpretation of the results at both a coarse- (general groups of triads) or fine-grained (individual colored triads) perspective. That said, we recognize that the interpretation is not straightforward and that this is a first effort at understanding these results, but we believe that having an algorithm to efficiently calculate the colored triad census will spur additional work towards interpreting and using the results. As a result better approaches therein will emerge with time and use.    \par
    
\section{Conclusions}

In this paper, we have extended the matrix algebra methods of \citet{Moody1998} to calculate the colored triad census for any network, directed or undirected, with an arbitrary number of colors in a relatively computationally efficient manner.  We have shown a number of mathematical results regarding the colored triad census, including a generalized equation for an arbitrary colored triad, the number of isomorphism classes for arbitrary numbers of colors, and the expectation and variances for colored triads.  We analyzed an empirical social network using our algorithm, and calculated approximate p-values for each colored triad, based on an analytic exact binomial test  for less complex null distributions, or approximately through simulation for more complex null distributions.  We have also shown the type of conclusions that can be drawn from these results, observing results that would not be feasible with many other currently available methods.  \par

One additional benefit of this method is that it can be directly used as a counting tool for sufficient statistics in network inference models, such as exponential random graphs (ERGM). The colored triad census essentially allows one to simultaneously evaluate the effect of local structure and node attribute on network structure in an ERGM, building off previous work where researchers explicated the ERGMs capacity for including the triad census \citep{Yaveroglu2015}.  We believe that the colored triad census is a useful technique with an efficient implementation that can be widely-applicable in social networks research, showing the continued importance of the triad census even in this era of stochastic models for complex networks. \par

\section{Acknowledgements}
    
The authors would like to thank the two anonymous reviewers and the editors for their contributions to this manuscript.  This work utilized the computational resources of the NIH HPC Biowulf cluster (http://hpc.nih.gov). This research was funded via National Human Genome Research Institute, National Institutes of Health (Grant number ZIA HG200335) and the Oxford Martin School, University of Oxford (Grant number LC1213-006).





\section{References}

\bibliographystyle{model1-num-names}
\bibliography{ColoredTriad.bib}







\clearpage

\appendix

\section{Variable and Functional Definitions}
\begin{table}[ht]
\centering
\begin{tabular}{l l}
\hline
\thead{Variable or function notation} & \thead{Description of variable or function}\\
\hline
$A$ & Adjacency matrix \\ 
$E$ & Symmetrized adjacency matrix\\ 
$\bar{E}$ & Complement of symmetrized adjacency matrix \\
$M$ & Adjacency matrix including only mutual ties \\ 
$C$ & Adjacency matrix including only asymmetric ties\\ 
$K^r$ & Coloring matrix for color $r$ \\
$R(i)$ & Function returning the color of node $i$ \\
$H(T,i,j)$ & Function returning the matrix of the edge in triad T \\
&between nodes i and j\\
$T$ & An arbitrary colored triad, with a MAN configuration, \\
&and colored triplet $r_1,r_2,r_3$\\
$L(T)$ & A function returning the number of unique colors for \\
&a given colored triad\\
$S(T,r)$ & Function returning the number of times color $r$ \\
&appears in colored triad $T$\\
$P(T)$ & The probability of observing triad $T$ \\
$E(T)$ & The expectation of triad $T$ under a binomial model \\
$V(T)$ & The variance of triad $T$ under a binomial model \\

\hline
\end{tabular}
\caption{List of variables, constants, and functions defined in this manuscript.}
\end{table}

\end{document}